\documentclass[aps,twocolumn]{revtex4}
\usepackage{graphicx}                       % Used for includegraphics
\usepackage{amsmath,amsfonts}
\usepackage{bm}                             % for bold math fonts
\graphicspath{{./Figures/}}                 % location for color  fig.

\newcommand{\noi}{\noindent}
\newcommand{\eq}{\begin{equation}}
\newcommand{\en}{\end{equation}}
\newcommand{\eqa}{\begin{eqnarray}}
\newcommand{\ena}{\end{eqnarray}}
\newcommand{\tr}{\mbox{Tr}}

\newcommand{\re}{\mbox{Re}}

\newcommand{\vb}{{\vec b}}

\newcommand{\vsigma}{{\vec\sigma}}

\newcommand{\Ra}{\Rightarrow}

\newcommand{\caa}{{\cal A}}
\def\be{\begin{equation}}
\def\ee{\end{equation}}
\def\bc{\begin{center}}
\def\ec{\end{center}}
\def\bea{\begin{eqnarray}}
\def\eea{\end{eqnarray}}

\begin{document}

% \rightline{HU-EP-03/79}

\title{On practical problems to compute the ghost propagator  \\
in $SU(2)$ lattice gauge theory}

\author{T.~Bakeev}
\email[]{bakeev@thsun1.jinr.ru}
\affiliation{Joint Institute for Nuclear Research, 141980 Dubna, Russia}

\author{E.--M.~Ilgenfritz}
\email[]{ilgenfri@physik.hu-berlin.de}
\affiliation{Institut f\"ur Physik, Humboldt Universit\"at zu Berlin,
Newton-Str. 15, D-12489 Berlin, Germany}

\author{V.~K.~Mitrjushkin}
\email[]{vmitr@thsun1.jinr.ru}
\affiliation{Joint Institute for Nuclear Research, 141980 Dubna, Russia \\
and 
Institute of Theoretical and Experimental Physics, 117259 Moscow, Russia}

\author{M.~M\"uller--Preussker}
\email[]{mmp@physik.hu-berlin.de}
\affiliation{Institut f\"ur Physik, Humboldt Universit\"at zu Berlin,
Newton-Str. 15, D-12489 Berlin, Germany}

\date{\today}

\begin{abstract}
In $SU(2)$ lattice pure gauge theory we study numerically the dependence 
of the ghost propagator $G(p)$ on the choice of Gribov copies in Lorentz 
(or Landau) gauge. We find that the effect of Gribov copies is
essential in the scaling window region, however, it tends to decrease
with increasing $\beta$. On the other hand, we find that at larger
$\beta$-values very strong fluctuations appear which can make
problematic the calculation of the ghost propagator.
\end{abstract}

% \pacs{12.38.Aw, 12.38.Gc, 11.15.Ha}
% \keywords{bbb}
\maketitle

% ----------------------------------
% Main Text
% ----------------------------------
\baselineskip=0.55cm

%%%%%%%%%%%%%%%%%%%%%%%%%%%%%%%%%%%%%%%%%%%%%%%%%%%%%%%%%%%%%%%%%%%%%%%%%%%%%%%
\section{Introduction}
\setcounter{equation}{0}

The nonperturbative study of the ghost propagator is of great interest
for the understanding of the mechanism of confinement. The Kugo-Ojima
confinement criterium \cite{Ojima,Kugo-Ojima} is formulated in terms of the
ghost propagator $G^{ab}(p)$ at $p \to 0$ and expresses the absence of
colored states from the spectrum of physical states. Moreover,
confinement of gluons has been related more directly to the suppression
of the gluon propagator in the limit $p \to 0$~\cite{Gribov}.
In both cases, the propagator in question is defined in the Landau
(or Lorentz) gauge.

\par
In a series of papers Zwanziger~\cite{Zwanziger-1} has suggested that this
behavior might result from the restriction of the fields in the transversal
plane $\Gamma=\{A:\partial_{\mu} A_{\mu}=0\}$ to the Gribov region
$\Omega=\{A:\partial_{\mu} A_{\mu}=0, M \ge 0\}$, where the Faddeev-Popov
operator $~M~$ is non-negative.

\par
From studies of the coupled Dyson-Schwinger equations for gluons and
ghosts~\cite{Smekal-Alkofer-Hauck,Alkofer-Smekal} it is well-known that the
infrared behaviour of gluon and ghost propagators is closely
related~\cite{Alkofer-Watson}: the gluon propagator
$D^{ab}_{\mu\nu}(p)=(\delta_{\mu\nu}-p_{\mu}p_{\nu}/p^2)~Z_{gl}(p^2)/p^2$
is damped in the infrared like $Z_{gl}(p^2) \propto (p^2)^{2\kappa}$,
while the ghost
propagator $G^{ab}(p)=\delta^{ab}~G(p)=\delta^{ab}~Z_{gh}(p^2)/p^2$ is more singular than the free
propagator, $Z_{gh}(p^2) \propto (p^2)^{-\kappa}$. In a particular truncation
scheme $\kappa = 0.595$ has been determined~\cite{IR-exponent,Zwanziger-4}.

\par
There are only relatively few previous lattice studies of the ghost
propagator~\cite{Suman-Schilling,Cucchieri-1,Nakajima-Furui,
Bloch-Cucchieri-Langfeld-Mendez},
in contrast to numerous investigations of the
gluon propagator~\cite{gluon-propagator-1,gluon-propagator-2,gluon-propagator-3}.
As for the latter, is not yet
clear from the lattice~\cite{open-question} whether
$Z_{gl}(p^2)/p^2 \to 0$ or $\ne 0$ with $p^2 \to 0$~\footnote{We notice that
the gluon propagator in the Gribov-copy free Laplacian gauge is finite in the
limit $p \to 0$, $V \to \infty$~\cite{Laplacian-gauge}.}.
The lattice volumes might still be insufficient to decide this question.
The singular behavior of $Z_{gh}(p^2)$ is seen to become stronger with
increasing volume~\cite{Cucchieri-1}. This supports the
expectation~\cite{Zwanziger-3}
that the sample of physically important gauge field configurations
$A \in \Gamma$, which constitutes the Euclidean functional integral,
in the thermodynamical limit $V \to \infty$ is concentrating towards the
edge of the Gribov region, the first Gribov horizon $\partial \Omega$
where the lowest non-vanishing eigenvalue of the Faddeev-Popov operator 
is approaching zero. This statement is the content of Zwanziger's horizon
condition~\cite{Zwanziger-3,Zwanziger-Schaden} which can be related to the
Kugo-Ojima criterion.

\par
All this is complicated by the non-uniqueness, first pointed out by
Gribov~\cite{Gribov}, of the intersection with $\Gamma$ of the gauge
orbit $A^{g}$ of any gauge field $A$, even if restricted to the Gribov
region $\Omega$. Practically, the Landau gauge is implemented by
maximizing (with respect to gauge transformations $g$) a certain gauge
functional. Usually, such a problem leads to more than a single maximum,
which are gauge copies (Gribov copies) of each other,
hence to a non-unique definition of gauge dependent observables.
Thus, in a lattice investigation one has to determine which observables
are really subject to the so-called Gribov problem which reflects
the dependence of an observable on the restriction (if possible) to
the copy corresponding to the absolute maximum of the gauge functional.
More precisely, one has to study whether this dependence disappears when
one is approaching the continuum and/or infinite volume limit.
Otherwise this would indicate the persistence of a real Gribov problem
to which Gribov has drawn the attention. On the lattice, the structure
of the Gribov region has been closer investigated under this aspect
only by Cucchieri~\cite{Cucchieri-2} some years ago.

\par
Here we are mainly dealing with the infrared behavior of the calculated
ghost propagator. In the result of a study for $SU(2)$ gluodyamics
\cite{Cucchieri-1}, Cucchieri came to the conclusion that the ghost propagator
depends on the selection of the highest among more and more maxima of the
gauge functional while the gluon propagator does not depend. This study
was restricted on one hand to the strong coupling region ($\beta =
0.0$, $0.8$, $1.6$) where these observations apply, and $\beta=2.7$
where no gauge copy dependence was seen at all. These $\beta$ values
are outside the physically interesting scaling region. In a more recent
paper~\cite{Bloch-Cucchieri-Langfeld-Mendez},
it has been reported that the gauge copy dependence
of the ghost propagator in the more interesting scaling region (at
$\beta = 2.15$, $2.2$, $2.3$ and $2.4$ for lattices $16^3 \times 32$)
has been found to be within the statistical errors,
on a level which is called Gribov noise.

\par
In the present paper we reanalyse the scaling region at $\beta = 2.2$,
$2.3$, $2.4$, $2.5$ and $2.6$ for lattices $8^4$ and $16^4$
by comparing two ensembles of gauge-fixed field configurations.
One ensemble (''fc'') consists of an arbitrary maximum (usually the first being
found), and the other consists of the best (relative) maximum (''bc'') among
$N_{copy}$ local maxima of the gauge functional. We find that the difference of
the ensemble averages of the ghost propagator for the lowest non-vanishing
lattice momentum between the two ensembles does not vanish, except for the
highest $\beta$ value. Hence the Gribov problem remains a serious obstacle
for a unique definition of the $SU(2)$ ghost propagator in the scaling region.
More serious is an unexpected observation in the higher-$\beta$ region.
We find an intermittent behavior of the ghost propagator estimator for
the lowest non-vanishing momentum, signalled by anomalously large,
isolated fluctuations of the ghost propagator $G(p_{min})$ (see below)
within the time history of uncorrelated configurations. We stress already
here that this behavior is {\it not} a Gribov copy problem since the
anomalous peaks of $G(p_{min})$ are observed both for the first and the
best Gribov copy, entering the ensembles ''fc'' and ''bc'', respectively. We have
tested whether this is correlated with various infrared observables.
For the time being, two hypothetic causes must be excluded as a viable
explanation of the phenomenon.

\par
In Section 2 we recall the definition of the gluon field $A_{\mu}$, the
definition of the Lorentz (or Landau) gauge, the structure of the
Faddeev-Popov operator and the definition of the ghost propagator. 
Details of the simulations, the gauge fixing and the observation of 
Gribov copies are reported in Section 3. In Section 4 we discuss 
the results on the ghost propagator. We conclude in Section 5.

%%%%%%%%%%%%%%%%%%%%%%%%%%%%%%%%%%%%%%%%%%%%%%%%%%%%%%%%%%%%%%%%%%%%%%%%%%%%%%%
\section{Faddeev-Popov operator and ghost propagator}
\setcounter{equation}{0}

\subsection{Definition of the gluon field and Faddeev-Popov operator}

For the Monte Carlo generation of ensembles of non-gauge-fixed
gauge field configurations we use the standard Wilson action~\cite{Wilson},
which for the case of an $SU(N)$ gauge group is written
\eqa
%% S(U) & = & \beta \sum_x\sum_{\mu >\nu}
S & = & \beta \sum_x\sum_{\mu >\nu}
\left[ 1 -\frac{1}{N}~\re~\tr \Bigl(U_{x\mu}U_{x+\mu;\nu}
U_{x+\nu;\mu}^{\dagger}U_{x\nu}^{\dagger} \Bigr)\right]\; ; \nonumber \\
& & \beta = 2N/g_0^2 \; .
\label{eq:action}
\ena
Here $g_0$ is a bare coupling constant and $U_{x\mu} \in SU(N)$ are
the link variables.
The field variables $U_{x\mu}$ transform as follows under gauge
transformations $g_x$ :
\eqa
U_{x\mu} &\stackrel{g}{\mapsto}& U_{x\mu}^{g}
= g_x^{\dagger} U_{x\mu} g_{x+\mu} \; ;
\qquad g_x \in SU(N) \; .
\label{eq:gaugetrafo}
\ena
For $SU(2)$ gauge links $U_{x\mu}$, a standard definition~\cite{Mandula-Ogilvie}
of the lattice gauge field (vector potential) $\caa_{x+\hat{\mu}/2,\mu}$ is
\eq
\caa_{x+\hat{\mu}/2,\mu} = \frac{1}{2i}~\Bigl( U_{x\mu}-U_{x\mu}^{\dagger}\Bigr)\; .
\label{eq:a_field}
\en
Therefore, for $SU(2)$, the link can be written
\eqa
U_{x\mu} & = & b^0_{x\mu}~{\hat 1} + i~\vb_{x\mu}~\vsigma
%% \equiv b^0_{x\mu}~{\hat 1} + i~\caa_{x+\hat{\mu}/2;\mu} \; ; \\
    = b^0_{x\mu}~{\hat 1} + i~\caa_{x+\hat{\mu}/2;\mu} \; ; \nonumber \\
b^0_{x\mu} & = & \frac{1}{2}~\tr~U_{x\mu} \; .
\label{eq:a_field_for_SU2}
\ena

\par
In lattice gauge theory the usual choice of the Landau gauge condition
is~\cite{Mandula-Ogilvie}
\eq
(\partial \caa)_{x} = \sum_{\mu=1}^4 \left( \caa_{x+\hat{\mu}/2;\mu}
- \caa_{x-\hat{\mu}/2;\mu} \right)  = 0 \; ,
\label{eq:diff_gaugecondition}
\en
which is equivalent to finding an extremum of the gauge functional
\eq
F_U(g) = \frac{1}{4V_4}~\sum_{x\mu}~\frac{1}{N}~\re~\tr~U^{g}_{x\mu} \;
\label{eq:gaugefunctional}
\en
with respect to gauge transformations $g_x~$. After replacing
$U \Rightarrow U^{g}$ at the extremum
the gauge condition (\ref{eq:diff_gaugecondition})
is satisfied. In what follows this gauge is referred to as Landau gauge.

\par
The lattice expression of the Faddeev-Popov operator $M^{ab}$
corresponding to $M^{ab} = - \partial_{\mu} D^{ab}_{\mu}$ in the
continuum theory
(where $D^{ab}_{\mu}$ is the covariant derivative in the adjoint
representation) is given by
\eqa
M^{ab}_{xy}  =  \sum_{\mu}~\Bigl\{ 
  \left( \bar{S}^{ab}_{x\mu} + \bar{S}^{ab}_{x-\hat{\mu};\mu}
\right)~\delta_{x;y} \Bigr. \nonumber     \\
\Bigl.   - \left( \bar{S}^{ab}_{x\mu} - \bar{A}^{ab}_{x\mu}
\right)~\delta_{y;x+\hat{\mu}} \Bigr.  \\
\Bigl.   - \left( \bar{S}^{ab}_{x-\hat{\mu};\mu}
+ \bar{A}^{ab}_{x-\hat{\mu};\mu} \right)~\delta_{y;x-\hat{\mu}} 
\Bigr\} \nonumber
\label{eq:M-form3}
\ena

\noi where

\eq
\bar{S}^{ab}_{x\mu} = \delta^{ab}~\frac{1}{2}~\tr~U_{x\mu} \; ;
\quad 
\bar{A}^{ab}_{x\mu} = -\frac{1}{2}~\epsilon^{abc}~A_{x+\hat{\mu}/2;\mu}^c \; .
\label{eq:abbreviations}
\en

\par
From the form (\ref{eq:M-form3}) it follows that
a trivial zero eigenvalue is always present, such that at the Gribov
horizon $\partial \Gamma$ the first  non-trivial zero eigenvalue
appears.  Conversely, it is easy to see that for constant field
configurations, with $b^{0}_{x\mu} = \bar{b}^{0}_{\mu}$ and
$b^{a}_{x\mu}=\bar{b}^{a}_{\mu}$ independent of $x$, there exist
eigenmodes of $M$ with a vanishing eigenvalue.
Thus, if the Landau gauge is properly implemented,
$M[U]$ is a symmetric and semi-positive definite matrix.

\subsection{Ghost propagator}

The ghost propagator $G^{ab}(x,y)$ is defined
as~\cite{Zwanziger-3,Suman-Schilling}
\be
G^{ab}(x,y) = \delta^{ab}~G(x-y) \equiv
\Bigl<\,\left(\, M^{- 1}\,\right)^{a\,b}_{x\, y} [U]\,\Bigl> \; ,
\label{ghostprop}
\ee
where $M[U]$ is
the Faddeev-Popov operator. Note that the ghost propagator becomes translational
invariant ({\it i.e.}, dependent only on $x-y$) and diagonal in color space only in
the result of averaging over the ensemble of gauge-fixed representants
(first or best gauge-fixed copies) of the original
Monte Carlo gauge configurations. The ghost propagator in momentum space can be
written as
\be
G(p)\, = \,\frac{1}{3 V} \sum_{x\mbox{,}\, y} e^{- 2 \pi i \, p \cdot (x - y)}
\Bigl<\,\left(\, M^{- 1}\,\right)^{a\,a}_{x\, y} [U]\,\Bigl> \; ,
\label{eq:ghostprop_in_momentumspace}
\ee
where $V=L^4$ is the lattice volume,
and the coefficient $\frac{1}{3V}$
is taken for a full normalization, including the indicated
color average over $a=1,..,3$.

\par
We mentioned above that $M[U]$ is a symmetric and semi-positive definite matrix.
In particular, this matrix is positive-definite in the subspace
orthogonal to constant vectors. The latter are zero modes of $M[U]$.
Therefore, it can be inverted by using a conjugate-gradient method, provided
that both the source $\psi^{a}(y)$ and the initial guess of the solution
are orthogonal to zero modes.
As the source we adopted the one proposed by Cucchieri
\cite{Cucchieri-1}:
\be
\psi^{a}(y) \,=\,\delta^{a c} \, e^{ 2 \pi i \, p \cdot y}
\qquad p \neq (0,0,0,0) \; ,
\label{eq:source}
\ee
for which the condition $\sum_{y}\, \psi^{a}(y)\,=\,0$ is
automatically imposed. Choosing the source in this way allows
to save computer time since,
instead of the summation over $x$ and $y$ in Eq.
(\ref{eq:ghostprop_in_momentumspace}), only the scalar product of $M^{-1}\psi$
with the source $\psi$ itself has to be evaluated.
In general, the gauge fixed configurations
can be used in a more efficient way when the inversion of $M$ is done on
sources for $c = 1,..,3$ such that the (adjoint) color averaging,
formally required in Eq. (\ref{eq:ghostprop_in_momentumspace}),
will be {\it explicitely} performed.

%%%%%%%%%%%%%%%%%%%%%%%%%%%%%%%%%%%%%%%%%%%%%%%%%%%%%%%%%%%%%%%%%%%%%%%%%%%%%%%
\section{Simulation details}

The numerical simulations have been done for $SU(2)$ pure gauge theory
using the standard Wilson action, for lattice volumes $L^4$ with
$L=8$ and $L=16$. At a given lattice size $L$ for each
$\beta$ value we have generated $N_{conf}$ independent mother configurations,
for which the Landau gauge was fixed $N_{copy}=20$ times, each time starting
from a random gauge transformation of the mother configuration, obtaining
in this way $N_{copy}$ Landau-gauge fixed copies.

\par
Two consecutive configurations (considered as independent) were separated
by 100 and 200 sweeps for lattice sizes $8^4$ and $16^4$, respectively.
Each sweep consisted of one local heatbath
update followed by 4 or 8 microcanonical updates~\cite{Adler} for $8^4$
or $16^4$ lattices. In all
our runs we have measured the integrated autocorrelation time for the
plaquette, for the Polyakov loop and for the ghost propagator (separately
for each momentum $p$). In all cases, the relation $\tau_{int} \sim 0.5$
was observed, showing that the consecutive configurations are effectively
independent.

\par
The actual measurements of the ghost propagator were done
for the ''first'', {\it i.e.} in fact an arbitrary gauge 
copy and for the ''best'' one among the $N_{copy}$ copies. 
If the first copy turned out to be the best, the ghost propagator was 
measured only once, and the result simultaneously entered the two different 
gauge-fixed ensemble averages. In the following the two ensembles are 
labelled ''fc'' and ''bc'',
refering to the first or the best gauge copy, respectively.
In Table \ref{tab:t1} we give, for each set of simulation parameters 
$(L,\beta)$, the number of times the first copy
produced turned {\it not} out to be the best, {\it i.e.}, did not
correspond to the relative maximum of $F_U(g_i)$ among the $N_{copy}$ copies.

\par
As the gauge fixing procedure we used
standard Los Alamos type overrelaxation with $\omega = 1.7$.
The iterations have been stopped when the following transversality
condition was satisfied:
\be
\max_{x\mbox{,}\, a} \, \Big|
\sum_{\mu=1}^4 \left( A_{x+\hat{\mu}/2;\mu}^a - A_{x-\hat{\mu}/2;\mu}^a \right)
\Big| \, < \, \epsilon_{lor} \; .
\label{eq:gaugefixstop}
\ee
We used the parameters $\epsilon_{lor}=10^{-10}$ or $10^{-9}$ 
for lattice size $8^4$ or $16^4$, respectively.
In our test runs it was found that further decreasing $\epsilon_{lor}$
does not affect the results for the ghost propagator. Also
it was checked that these values of $\epsilon_{lor}$  are
sufficient for identifying, according only to the values of $F_U(g_i)$,
Gribov copies which are actually global gauge transformations of each
other and conversely for distinguishing this from the case of
actually inequivalent lattice Gribov copies.

\par
In Table \ref{tab:t1},
for each set of simulation parameters $(L,\beta)$,
we present also the number of configurations for which Gribov
copies have been found and the total number of different
Gribov copies.

\par
The momenta $p$ for the propagator $G(p)$ were taken with
all spatial components put equal to zero:
$p = (0,0,0,k_4/L)$, where $k_4$ was restricted to $k_{4}=1,2,3,4$.

%%%%%%%%%%   \normalsize
%%%%%%%%%%%%%%%%%%%%%%%%%%%%%%%%%%%%%%%%%%%%%%%%%%%%%%%%%%%%%%%%%%
%%%%%%%%%%%%%%%%%%%%%%%%%%% Table 1 %%%%%%%%%%%%%%%%%%%%%%%%%%%%%%
%%%%%%%%%%%%%%%%%%%%%%%%%%%%%%%%%%%%%%%%%%%%%%%%%%%%%%%%%%%%%%%%%%
%MMP? notation changed acc. to the referee's remark:
%EMI? in addition wording changed a bit
\begin{table}
\begin{center} \begin{tabular}{||c|c|c|c|c||}
\hline size & $\beta$ & $N^{conf}_{multiple~copies} / N^{conf}$&
$N^{copies}_{nonequiv} / N^{copies}_{total}$ & $N_{fc \ne bc}$ \\
\hline\hline
$8^4$  &1.6 &500 / 500& 8263 / 10000 & 446\\
       &2.0 &490 / 500& 4431 / 10000 & 354\\
       &2.1 &468 / 500& 3460 / 10000 & 311\\
       &2.2 &426 / 500& 2180 / 10000 & 235\\
       &2.3 &301 / 500& 1364 / 10000 & 150\\
       &2.4 &184 / 500& ~877 / 10000 & 92 \\
\hline\hline
$16^4$ &2.0 &25 / 25 & 500 / 500  & 25 \\
       &2.1 &25 / 25 & 500 / 500  & 25 \\
       &2.2 &25 / 25 & 500 / 500  & 23 \\
       &2.3 &25 / 25 & 494 / 500  & 25 \\
       &2.4 &25 / 25 & 337 / 500  & 23 \\
       &2.5 &24 / 25 & 169 / 500  & 14 \\
\hline\hline
\end{tabular}
\end{center}
\caption{\label{tab:t1}
%%%%%%%%%%%%%%%%%   \footnotesize
The Table shows in the 3rd column the number $N^{conf}_{multiple~copies}$
of configurations, for which non-equivalent Gribov copies have actually 
been obtained, out of a total number $N^{conf}$ of configurations which 
underwent gauge fixing; in the 4th column the total number 
$N^{copies}_{nonequiv}$ of non-equivalent Gribov copies out of a 
total number $N^{copies}_{total} = N^{conf} \times N_{copy}$ of gauge 
copies under investigation. The last column presents the number of times 
out of $N^{conf}$ that the first copy was not identical to the best 
(relative maximum) copy. }
\end{table}
%MMP! EMI!

%%%%%%%%%%%%%%%%  \normalsize

%%%%%%%%%%%%%%%%%%%%%%%%%%%%%%%%%%%%%%%%%%%%%%%%%%%%%%%%%%%%%%%%%%
%%%%%%%%%%%%%%%%%%%%%%%%%%% Table 2 %%%%%%%%%%%%%%%%%%%%%%%%%%%%%%
%%%%%%%%%%%%%%%%%%%%%%%%%%%%%%%%%%%%%%%%%%%%%%%%%%%%%%%%%%%%%%%%%%

\begin{table}
\vspace*{-0.7cm}
\protect\small
\begin{center}
\begin{tabular}{||c|c|c|c|c|c|c||}
\hline\hline
\multicolumn{7}{|c|}{$8^4$ lattice}\\ \hline\hline
$\beta$ & $n_{meas}$ & Copy &
$ k_4=1 $ & $ k_4=2 $ & $ k_4=3 $ & $ k_4=4$
\\ \hline\hline
 $ 1.6 $ & $500$ & bc & $ 6.58 (4)
 $ & $ 1.327(5)
 $ & $ 0.628(2) $ & $ 0.501(1) $ \\ \hline
 $ 1.6 $ & $500$ & fc & $ 7.02(6)
 $ & $ 1.363(5)
 $ & $ 0.638(1) $ & $ 0.508(1) $ \\ \hline \hline
 $ 2.0 $ & $500$ & bc & $ 5.15(3)
 $ & $ 1.013(2)
 $ & $ 0.491(1)
 $ & $ 0.3970(4) $ \\ \hline
 $ 2.0 $ & $500$ & fc & $ 5.46(9)
 $ & $ 1.028(3)
 $ & $ 0.495(1)
 $ & $ 0.3995(6) $ \\ \hline \hline
 $ 2.1 $ & $500$ & bc & $ 4.62(3)
 $ & $ 0.920(2)
 $ & $ 0.4545(6)
 $ & $ 0.3701(4)
 $ \\ \hline
 $ 2.1 $ & $500$ & fc & $ 4.89(7)
 $ & $ 0.935(3)
 $ & $ 0.4573(8)
 $ & $ 0.3719(5)
 $ \\ \hline \hline
 $ 2.2 $ & $500$ & bc & $ 4.06(3)
 $ & $ 0.823(2) $ & $0.4189(4) $ & $ 0.3444(3) $ \\ \hline
 $ 2.2 $ & $500$ & fc & $ 4.26(4)
 $ & $ 0.833(2) $ & $ 0.4203(5) $ & $ 0.3450(3) $ \\
       \hline\hline
 $ 2.3 $ & $500$ & bc & $ 3.60(4)
 $ & $ 0.744(1)
 $ & $ 0.3903(3) $ & $ 0.3238(2) $ \\ \hline
 $ 2.3 $ & $500$ & fc & $ 3.65(4)
 $ & $ 0.747(2)
 $ & $ 0.3909(4) $ & $ 0.3241(2) $ \\ \hline \hline
 $ 2.4 $ & $500$ & bc & $ 3.38(5)
 $ & $ 0.691(1)
 $ & $ 0.3710(4)
 $ & $ 0.3098(2) $ \\ \hline
 $ 2.4 $ & $500$ & fc & $ 3.47(7)
 $ & $ 0.692(2)
 $ & $ 0.3712(4)
 $ & $ 0.3099(2)
 $ \\ \hline \hline
%
%\end{tabular}
%\end{center}
%
%\normalsize
%\vspace*{-0.7cm}
%\protect\small
%\begin{center}
%\begin{tabular}{||c|c|c|c|c|c|c||}
\multicolumn{7}{|c|}{$16^4$ lattice}\\ \hline\hline
$ \beta $ & $n_{meas}$ & Copy &
$ k_4=1 $ & $ k_4=2 $ & $ k_4=3 $ & $ k_4=4$
\\ \hline\hline
$2.2$ & $296$ &bc& $20.1(1)$ & $3.87(1)$ & $1.494(2)$ & $0.8078(6)$
\\ \hline
$2.2$ & $296$ &fc& $21.3(1)$ & $3.97(1)$ & $1.509(2)$ & $0.8115(6)$
\\ \hline \hline
$2.3$ & $270$ &bc& $17.3(1)$ & $3.29(1)$ & $1.303(1)$ & $0.7248(4)$
\\ \hline
$2.3$ & $270$ &fc& $18.0(1)$ & $3.33(1)$ & $1.310(2)$ & $0.7268(5)$
\\ \hline \hline
$2.4$ & $370$ &bc& $14.8(1)$ & $2.83(1)$ & $1.165(1)$ & $0.6673(3)$
\\ \hline
$2.4$ & $370$ &fc& $15.6(1)$ & $2.87(1)$ & $1.171(1)$ & $0.6690(3)$
\\ \hline \hline
$2.5$ & $294$ &bc& $13.7(2)$ & $2.56(1)$ & $1.088(1)$ & $0.6353(3)$
\\ \hline
$2.5$ & $294$ &fc& $13.9(2)$ & $2.58(1)$ & $1.090(1)$ & $0.6358(3)$
\\ \hline \hline
$2.6$ & $229$ &bc& $13.6(4)$ & $2.41(1)$ & $1.043(2)$ & $0.6161(5)$
\\ \hline
$2.6$ & $229$ &fc& $13.8(4)$ & $2.41(1)$ & $1.044(2)$ & $0.6164(5)$
\\ \hline \hline
\end{tabular}
\end{center}
\caption{\label{tab:t2}
%%%%%%%%%%%%%   \footnotesize
The ghost propagator $G(p)$ from Eq. (\ref{eq:ghostprop_in_momentumspace})
as a function of $k_{4}=1,2,3,4$.
We have set the momentum $p = (0,0,0,k_4/L)$, where $L=8,16$
is the lattice size. The averages over the gauge
configurations in Eq. (\ref{eq:ghostprop_in_momentumspace}) were
taken in two different ways: ''fc'' means the average taking
only the gauge-fixed copy generated first for each
configuration, ''bc'' means the average over only the
best (relative maximum) copy among $20$ different
gauge-fixed copies that we have generated.
}
\end{table}
%%%%%%  \normalsize

%%%%%%%%%%%%%%%%%%%%%%%%%%%%%%%%%%%%%%%%%%%%%%%%%%%%%%%%%%%%%%%%%%%%%%%%%%%%%%%
\section{Discussion of the Results}
\setcounter{equation}{0}

From Table \ref{tab:t1} one can learn that at the lattice size $8^4$ the 
fraction of Monte Carlo configurations which are represented by more than one
gauge-fixed configurations (among 20 attempts to find copies) drastically
begins to decrease at $\beta=2.3$. Parallel to this also the multiplicity
of {\it actually different} copies among 20 drops down.
The decrease of the number of available {\it basins of attraction} for the
gauge fixing process is a finite-volume effect.
For the bigger lattice size ($16^4$) one sees that the fraction of Monte
Carlo configurations with more than one gauge-fixed configurations
practically does not depend on $\beta$.
However, the multiplicity of non-equivalent copies among the 20 obtained
copies starts to decrease from $\beta=2.3$.

\par
From Table \ref{tab:t2} one can see for separate small momenta, how the average
value of the ghost propagator differs between the two ways to deal with
the Gribov copy problem: to ignore it ($N_{copy}=1$) or to inspect 
$N_{copy}=20$ copies. 
%MMP? 
%EMI? wording changed a bit  
For all momenta, the ensemble consisting of the first copies ('fc')
%% values ('fc') 
turns out to give slightly larger values 
than the ensemble including always the best copy ('bc').
%% best copy ones ('bc'). 
%%%%% FIGURE1
\begin{figure*}[t]
\centering
\includegraphics[width=10cm]{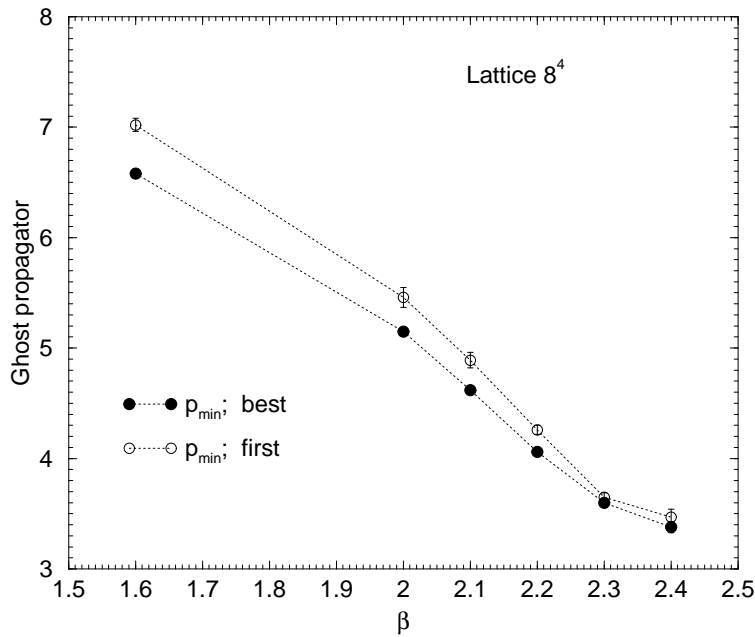}
\caption{The $\beta$ dependence and Gribov copy sensitivity
of the ghost propagator $G(p_{min})$ at minimal momentum 
$p_{min}$ on the $8^4$ lattice. Filled symbols correspond 
to the 'bc' ensemble, open symbols to the 'fc' ensemble 
(see the text).}
\label{fig:gho_08x08_m1}
\end{figure*}
For the lowest non-vanishing momentum  
this is shown in Fig. \ref{fig:gho_08x08_m1} for the lattice $8^4$.
It is visible that for $\beta \in [1.5,2.4]$ the difference of $G(p_{min})$
between the two ways of averaging is clearly outside the statistical error.

\par
\begin{figure*}
\centering
\includegraphics[width=10cm]{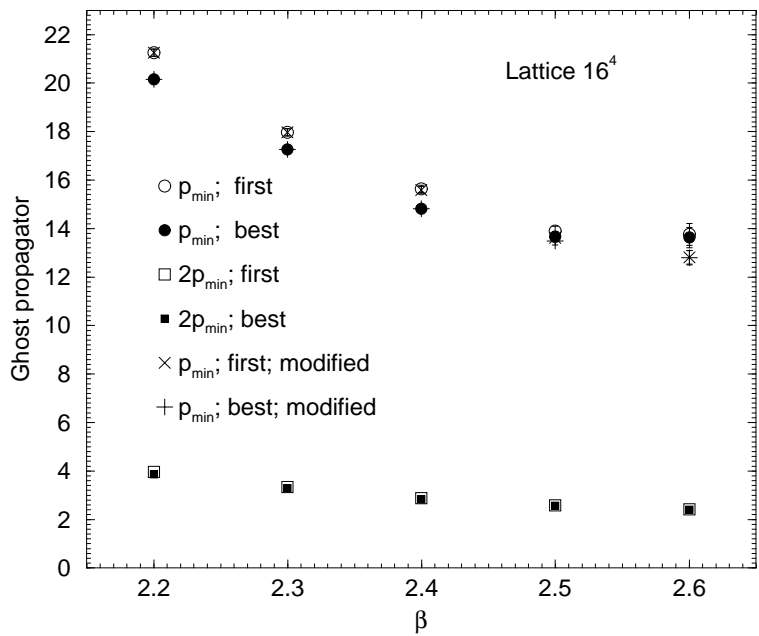}
\caption{The $\beta$ dependence and Gribov copy sensitivity
of the ghost propagator $G(p)$ at minimal momentum $p_{min}$ 
and next-to-minimum momentum $2p_{min}$ on the $16^4$ lattice. 
Filled symbols correspond to the 'bc' ensemble, open symbols 
to the 'fc' ensemble. At $\beta=2.6$ the minimal momentum 
ghost propagator is sensitive to the inclusion or exclusion 
of "exceptional configurations" (see text).}
\label{fig:gho_16x16_m1m2_mod}
\end{figure*}

In Fig. \ref{fig:gho_16x16_m1m2_mod}, for the bigger lattice $16^4$,
the ghost propagator 
%% added
values for the two lowest momenta are compared with 
respect to the dependence on Gribov copies for $\beta \in [2.2,2.6]$.
Whereas for the lowest momentum the results resemble those of the
smaller lattice, for the second lowest momentum they are practically 
indistinguishable at the given scale. For increasing $\beta$ the
difference becomes of the order of the statistical error (Gribov noise).
At $\beta=2.6$ the ghost propagator data even for the {\it lowest} momentum
fall together within error bars.
%MMP! EMI!
This indicates that the Gribov problem has disappeared for the
ghost propagator there.
%%%%% FIGURE2

\par
\begin{figure*}
\centering
\includegraphics[width=10cm]{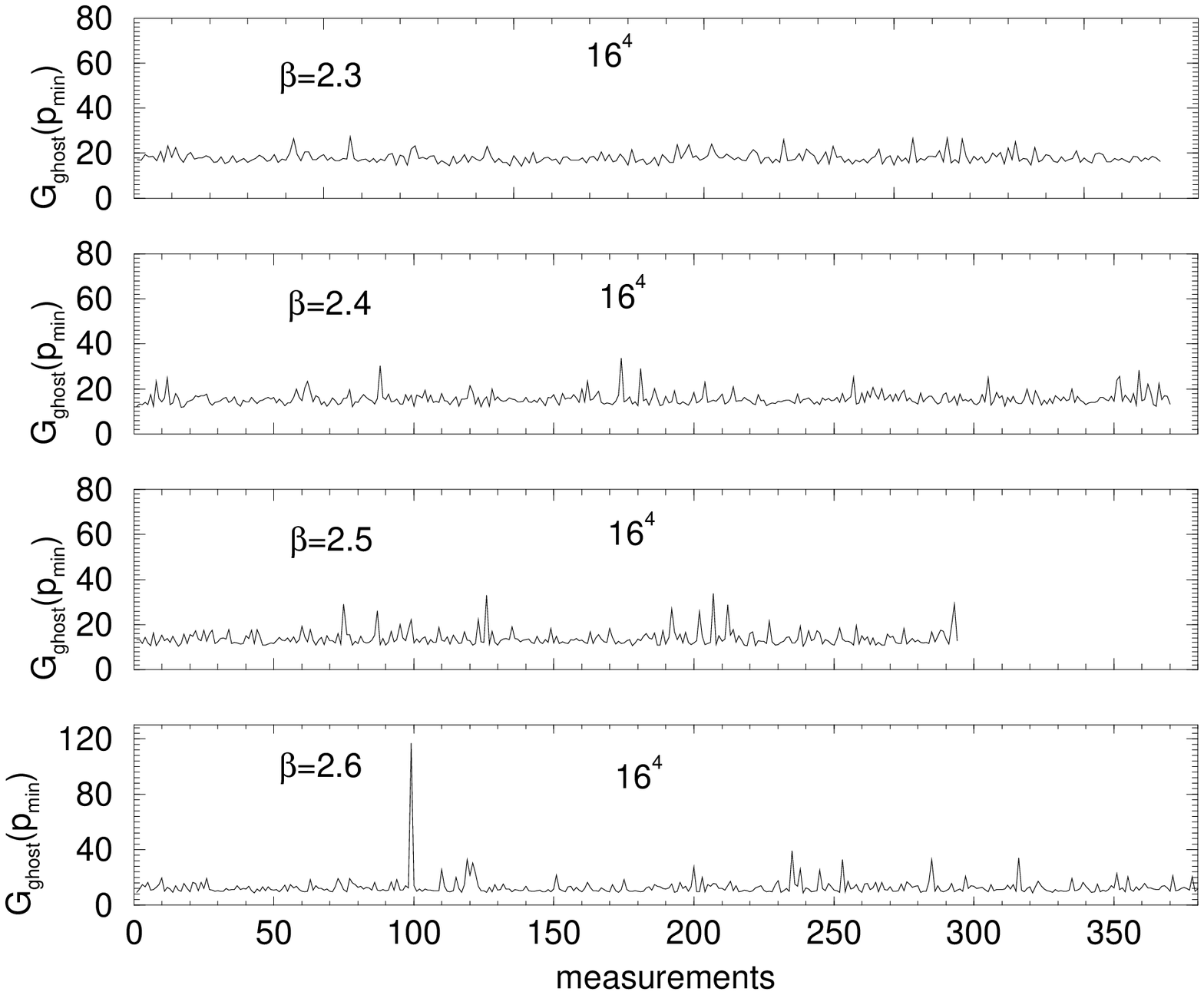}
\caption{Monte Carlo time histories of the ghost propagator 
$G(p_{min})$ for various $\beta$ on the $16^4$ lattice. 
The frequency of the occurrence of ''exceptional configurations'' 
increases with higher $\beta$.}
\label{fig:his_16x16_m1}
\end{figure*}

Instead, at $\beta=2.6$ a new problem arises
which can be recognized already in Fig. \ref{fig:gho_16x16_m1m2_mod}
where we also demonstrate
how, at $\beta=2.6$, the average for the ghost propagator at the lowest momentum
would be influenced by the removal
of ''exceptional configurations''. These are signalled as spikes in the Monte Carlo
time histories of the corresponding observable shown in Fig. \ref{fig:his_16x16_m1}
for $\beta=2.6$. Precursors of this phenomenon are visible there at lower
$\beta$, too, but for $\beta=2.6$ the effect becomes notable.
We notice that these spikes occur in the first as well as in the best
gauge-fixed copy.  Therefore, the existence of these ''exceptional configurations''
is definitely {\it not} a result of gauge fixing.
%%%%% FIGURE3

\par
In order to explore what the essence of these ''exceptional configurations'' is,
we have looked for correlations with certain ''toron'' excitations on one
hand and with different Polyakov loops on the other.

\par
\begin{figure*}
\centering
\includegraphics[width=10cm]{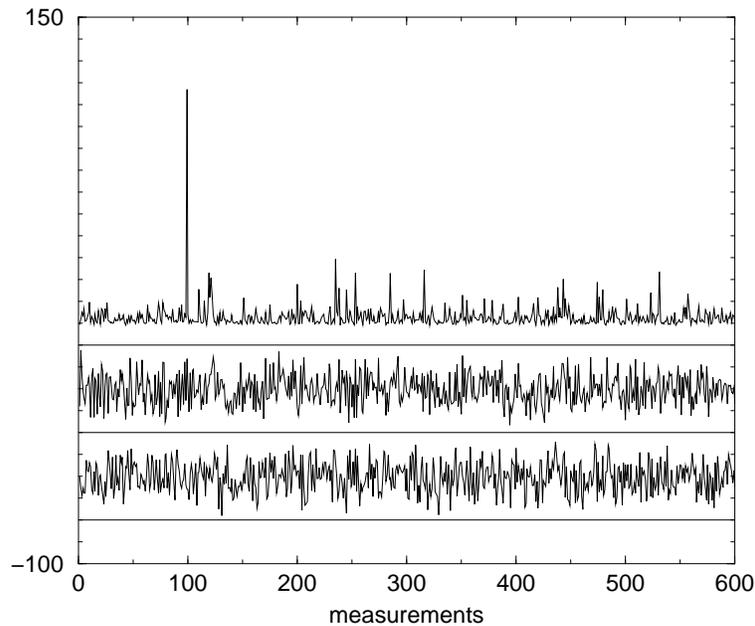}
\caption{Monte Carlo time histories of the ghost propagator $G(p_{min})$
(above) for $\beta=2.6$ on the $16^4$ lattice, compared with the histories 
of the toron fields $T_{4}$ (middle) and $T_{1}$ (below). The fluctuations
of the latter have been arbitrarily rescaled and shifted for better visual 
inspection.}
\label{fig:ghost-toron}
\end{figure*}

In the first case we followed the procedure applied by Kovacs \cite{Kovacs} 
for extracting the toron content of Monte Carlo gauge field 
configurations~\footnote{In an attempt to reconstruct hadronic correlators
from model configurations derived from lattice Monte Carlo configurations
he found it necessary to augment the instanton content of the latter - as
extracted via smoothing - by an appropriate ''toron'' field extracted as
we explain in the text. Indeed, this mixture turned out essential to reproduce
mesonic correlators in his ''instanton plus toron'' model of the vacuum.}.
We evaluated for all four directions $\mu$ on the lattice the corresponding
holonomies over a $\mu$-slice fixed at $x_{\mu}=1$ 
\be
{\cal P}_{\mu}(x) = \prod_{s=0}^{L-1} U_{x+s\hat{\mu};\mu} \; .
\label{eq:holonomy}
\ee
We averaged this quantity over the $\mu$-slice,
\be
\bar{{\cal P}}_{\mu} = \frac{1}{L^3} \sum_{x;x_{\mu}=1} {\cal P}_{\mu}(x) \; .
\label{eq:average_holonomy}
\ee
These {\it gauge dependent} quantities were normalized to $SU(2)$ in the usual
way
\be
\bar{{\cal P}}_{\mu} \Ra \bar{{\cal P}}_{\mu}/
\sqrt{\mathrm{det} \bar{{\cal P}}_{\mu}} \; .
\label{eq:projected_average_holonomy}
\ee
Then the anticipated homogeneous toron field is given by links $\bar{U}_{x\mu}$
independent of $x$, which are required to reproduce
$\bar{{\cal P}}_{\mu}$ as follows :
\be
(\bar{U}_{x\mu})^L = \bar{{\cal P}}_{\mu} \; .
\label{eq:U_toron}
\ee
The corresponding toron gluon field can be extracted as
\be
\caa^{toron}_{x+\hat{\mu}/2;\mu} =
\frac{1}{2i}~\Bigl( \bar{U}_{x\mu}-\bar{U}_{x\mu}^{\dagger} \Bigr)\; .
\label{eq:A_toron}
\ee
We have plotted the time history of the lowest momentum ghost propagator
together with the toron observable
\be
T_{\mu} = \sum_{a=1}^{3} \tr~({\caa}^{toron}_{x+\hat{\mu}/2;\mu})^2 \; ,
\label{eq:toron_signal}
\ee
\noi defined separately for the four Euclidean directions. We noticed
that the previously mentioned spikes ("exceptional configurations")
occur independent of spikes of this toron observable in each of the
Euclidean directions.
%MMP? EMI ?
We demonstrate this in Fig. \ref{fig:ghost-toron} which shows the Monte Carlo
history of the lowest-momentum ghost propagator (upper panel) together with 
the histories of the toron fields $T_{\mu}$ for $\mu=4$ (middle) and $\mu=1$ 
(lower panel).
%MMP! EMI !
%%%%% FIGURE4

\par
We also checked the Monte Carlo sample for eventual correlations with
the average Polyakov loop

\be
L_{\mu} = \frac{1}{2} \tr~\bar{{\cal P}}_{\mu} \; .
\label{eq:average_Polyakov_loop}
\ee

\begin{figure*}
\centering
\includegraphics[width=10cm]{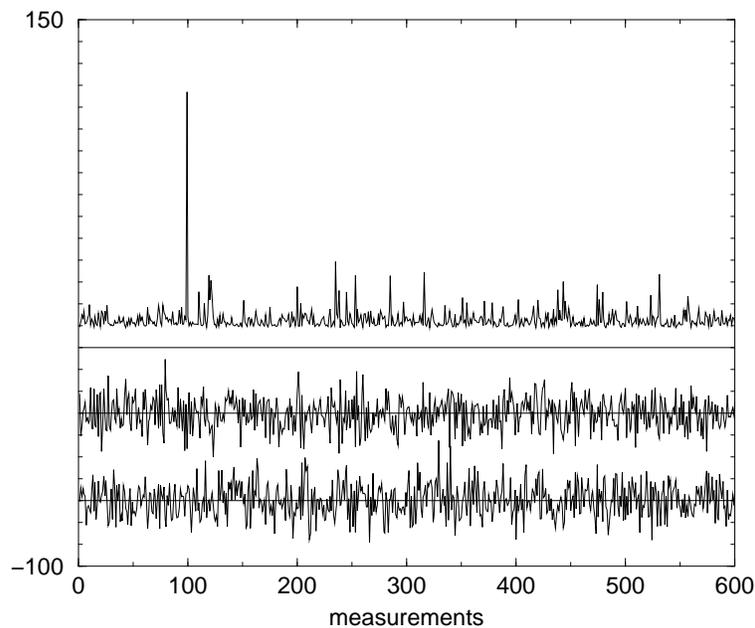}
\caption{Monte Carlo time histories of the ghost propagator $G(p_{min})$
(above) for $\beta=2.6$ on the $16^4$ lattice, compared with the histories 
of the average Polyakov lines $L_{4}$ (middle) and $L_{1}$ (below). The 
fluctuations of the latter have been arbitrarily rescaled and shifted for 
better visual inspection.}
\label{fig:ghost-polyak}
\end{figure*}

\noi 
Similarly, 
%MMP? EMI ?
we illustrate in Fig. \ref{fig:ghost-polyak} that there are no correlations 
between the spikes of the lowest momentum ghost propagator with extremal 
fluctuations of the average Polyakov loop in any of the four directions. 
Shown in the Fig. \ref{fig:ghost-polyak} are, beside the history of the 
lowest-momentum ghost propagator (upper panel), the histories of the 
average Polyakov lines $L_{\mu}$ for $\mu=4$ (middle) and $\mu=1$ 
(lower panel).
%MMP! EMI !
%%%%% FIGURE5

%%%%%%%%%%%%%%%%%%%%%%%%%%%%%%%%%%%%%%%%%%%%%%%%%%%%%%%%%%%%%%%%%%%%%%%%%%%%%%%
\section{Conclusions}
\setcounter{equation}{0}

In this work we studied numerically the dependence of the ghost
propagator $G(p)$ in pure gauge $SU(2)$ theory on the choice of Gribov
copies in Lorentz (or Landau) gauge with the special focus on the
physically interesting scaling region. All simulations have been
performed on the $~8^4~$ and $~16^4~$ lattices.

\par
We found  that the effect of Gribov copies
is essential in the
scaling window region. Therefore, the Gribov problem remains a serious
obstacle for a unique definition of the $SU(2)$ ghost propagator in the
scaling region. However, it tends to decrease with increasing
$\beta$ values.

\par
Another -- and more serious -- problem is
presented by the unexpected observation, in the higher-$\beta$ region,
of anomalously large, isolated fluctuations
of the ghost propagator $G(p_{min})$  within the time history of
uncorrelated configurations. These strong fluctuations
make problematic the calculation of the ghost propagator.

\par
We believe that this problem deserves a more thorough study,
in particular how to interpret the relevant configurations.
If there is nothing physically wrong with them, much more
statistics is necessary to get a reliable result.

%%%%%%%%%%%%%%%%%%%%%%%%%%%%%%%%%%%%%%%%%%%%%%%%%%%%%%%%%%%%%%%%%%%%%%%%%%%%%%%
\subsection*{Acknowledgment}

This work has been supported by the grant INTAS--00--00111 and RFBR
grants 02--02--17308 and 99--01--00190. One of us (T.~B.) recognizes
the support of young scientist's grant 02-01-06064.
E.--M.~I. is supported by DFG through the DFG-Forschergruppe 
''Lattice Hadron Phenomenology'' (FOR 465).

%%%%%%%%%%%%%%%%%%%%%%%%%%%%%%%%%%%%%%%%%%%%%%
% \bibliographystyle{h-physrev3}


\begin{thebibliography}{99}
%%    {10}
%% \itemsep=0.2cm

\bibitem{Ojima} 
I.~Ojima, 
\newblock Nucl. Phys. {\bf B 143}, 340 (1978).

\bibitem{Kugo-Ojima} 
T.~Kugo and I.~Ojima, 
\newblock Prog. Theor. Phys. Suppl. {\bf 66}, 1 (1979).

\bibitem{Gribov} 
V.~N.~Gribov, 
\newblock Nucl. Phys. {\bf B 139}, 1 (1978).

\bibitem{Zwanziger-1} 
D.~Zwanziger, 
\newblock Nucl. Phys. {\bf B 364}, 127 (1991); Nucl. Phys. {\bf B 378}, 525 (1992).

\bibitem{Smekal-Alkofer-Hauck} 
L.~von~Smekal, R.~Alkofer and A.~Hauck,
\newblock Phys. Rev. Lett. {\bf 79}, 3591 (1997);  
L.~von~Smekal, A.~Hauck and R.~Alkofer, 
\newblock Ann. Phys. (NY) {\bf 267}, 1 (1998), Erratum-ibid. {\bf 269},182 (1998).

\bibitem{Alkofer-Smekal} 
R.~Alkofer and L.~von~Smekal, 
\newblock Phys. Rept. {\bf 353}, 281 (2001).

\bibitem{Alkofer-Watson} 
P.~Watson and R.~Alkofer, 
\newblock Phys. Rev. Lett. {\bf 86}, 5239 (2001).

\bibitem{IR-exponent}
R.~Alkofer, C.S.~Fischer and L.~von~Smekal, 
\newblock Eur. Phys. J. {\bf A 17}, 773 (2003);
Prog. Part. Nucl. Phys. {\bf 50}, 317 (2003), 
e-print archive: nucl-th/0301048.

\bibitem{Zwanziger-4} 
D.~Zwanziger, 
\newblock Phys. Rev. {\bf D 65}, 094039 (2002); 
Phys. Rev. {\bf D 67}, 105001 (2003).

\bibitem{Suman-Schilling} 
H.~Suman and K.~Schilling, 
\newblock Phys. Lett. {\bf B 373}, 314 (1996).

\bibitem{Cucchieri-1} 
A.~Cucchieri, 
\newblock Nucl. Phys. {\bf B 508}, 353 (1997).

\bibitem{Nakajima-Furui}
H.~Nakajima and S.~Furui, 
\newblock Nucl. Phys. {\bf B} (Proc. Suppl.) {\bf  73}, 865 (1999);
          Nucl. Phys. {\bf B} (Proc. Suppl.) {\bf  83}, 521 (2000);
          Nucl. Phys. {\bf A 680}, 151 (2000);
H.~Nakajima, S.~Furui and A.~Yamaguchi, 
\newblock Nucl. Phys. {\bf B} (Proc. Suppl.) {\bf  94}, 558 (2001);
H.~Nakajima and S.~Furui, 
\newblock Nucl. Phys. {\bf B} (Proc. Suppl.) {\bf 119}, 730 (2003).

\bibitem{Bloch-Cucchieri-Langfeld-Mendez} 
J.C.R~Bloch, A.~Cucchieri, K.~Langfeld and T.~Mendez, 
\newblock Nucl. Phys. {\bf B} (Proc. Suppl.) {\bf 119}, 736 (2003); \\
K.~Langfeld, J.C.R.~Bloch, J.~Gattnar, H.~Reinhardt, A.~Cucchieri 
and T.~Mendes,
\newblock talk given at {\em 5th International Conference on Quark 
Confinement and the Hadron Spectrum, Gargnano, Brescia, Italy, 
10-14 Sep 2002}, edited by N. Brambilla, G. M. Prosperi; 
River Edge, N.J., World Scientific (2003), pp. 297-299;
e-print archive: hep-th/0209173. 

\bibitem{gluon-propagator-1}
J.~E.~Mandula, 
\newblock Phys. Rept. {\bf 315}, 273 (1999).

\bibitem{gluon-propagator-2}
G.~Damm, W.~Kerler and V.K.~Mitrjushkin, 
\newblock Phys. Lett. {\bf B 433}, 88 (1998); \\
A.~Cucchieri, 
\newblock Phys. Rev. {\bf D 60}, 034508 (1999); \\
K.~Langfeld, H.~Reinhardt and J.~Gattnar, 
\newblock Nucl. Phys. {\bf B 621}, 131 (2002); \\
A.~Cucchieri, T.~Mendes and A.~R.~Taurines, 
\newblock Phus. Rev. {\bf D 67}, 091502 (2003).

\bibitem{gluon-propagator-3}
D.~B.~Leinweber et al. (UKQCD collaboration), 
\newblock Phys. Rev. {\bf D 58}, 031501 (1988);
          Phys. Rev. {\bf D 60}, 094507 (1999), 
       Erratum-ibid. {\bf D 61}, 079901 (2000); \\
H.~Nakajima and S.~Furui, 
\newblock Nucl. Phys. {\bf B} (Proc. Suppl.) {\bf  73}, 635 (1999); \\
F.D.R.~Bonnet, P.~O.~Bowman, D.~B.~Leinweber and A.~G.~Williams, 
\newblock Phys. Rev. {\bf D 62}, 051501 (2000);
F.D.R.~Bonnet, P.~O.~Bowman, D.~B.~Leinweber, A.~G.~Williams 
and J.~M.~Zanotti,
\newblock Phys. Rev. {\bf D 64}, 034501 (2001).

\bibitem{open-question}
see the last two of Ref.~\cite{gluon-propagator-3}

\bibitem{Laplacian-gauge}
C.~Alexandrou, P.~de~Forcrand and E.~Follana, 
\newblock Phys. Rev. {\bf D 65}, 114508 (2002); \\
P.~O.~Bowman, U.~M.~Heller, D.~B.~Leinweber and A.~G.~Williams, 
\newblock Phys. Rev. {\bf D 66}, 074505 (2002).

\bibitem{Zwanziger-3} 
D.~Zwanziger, 
\newblock Nucl. Phys. {\bf B 412}, 657 (1994).

\bibitem{Zwanziger-Schaden}
M.~Schaden and D.~Zwanziger,
\newblock in {\em Proceedings of the Workshop on Quantum Infrared Physics, 
Paris, France, 6-10 Jun 1994}, edited by H.M. Fried and B. Muller; 
River Edge, N.J., World Scientific (1995), pp. 0010-17;    
e-print archive: hep-th/9410019.

\bibitem{Cucchieri-2} 
A.~Cucchieri, 
\newblock Nucl. Phys. {\bf B 521}, 365 (1998).

\bibitem{Wilson} 
K.~Wilson, 
\newblock Phys. Rev. {\bf D 10}, 2445 (1974).
.
\bibitem{Mandula-Ogilvie} 
J.~E.~Mandula and M.~Ogilvie, 
\newblock Phys. Lett. {\bf B 185}, 127 (1987).

\bibitem{Adler} 
S.~L.~Adler, 
\newblock Nucl. Phys. {\bf B} (Proc. Suppl.) {\bf 9}, 437 (1989).

\bibitem{Kovacs} 
T.~G.~Kovacs, 
\newblock Phys. Rev. {\bf D 62}, 034502 (2000).

\end{thebibliography}
\end{document}